\documentstyle[12pt,times,aaspp4]{article}
  
\newcommand\mathmode[1]{\ifmmode {#1} \else {$#1\mkern-5mu$} \fi}
\newcommand\wrange[2]{\mathmode{ {#1} < \lambda < {#2} \: {\rm \AA}}}
\newcommand\wless[1]{\hbox{\mathmode{\lambda < {#1} \: {\rm \AA}}}}

\begin{document}

\title{The G Dwarf Problem Exists in Other Galaxies}
  
\author{Guy Worthey\altaffilmark{1}}  
\affil{Department of Astronomy, University of Michigan, Ann Arbor, MI  
48109-1090}
\authoremail{worthey@astro.lsa.umich.edu}

\author{ Ben Dorman\altaffilmark{2}}  
\affil{Laboratory for Astronomy \& Solar Physics, 
NASA/Goddard Space Flight Center, Greenbelt, MD 20771} 
\authoremail{Ben.Dorman@gsfc.nasa.gov}

\author{ Lewis A. Jones}
\affil{Department of Physics and Astronomy, University of North Carolina, CB\#3255 Phillips Hall, Chapel Hill, NC 27599-3255}
\authoremail{lewis@physics.unc.edu}

\altaffiltext{1}{Hubble Fellow} 
\altaffiltext{2}{NAS/NRC Resident Research Associate}  
\begin{abstract}

Stellar population models with abundance distributions determined from
the analytic Simple model of chemical evolution fail to match
observations of the nuclei of bulge-dominated galaxies in three
respects. First, the spectral energy distribution in the
mid-ultraviolet range \wrange{2000}{2400} exceeds observation by $\sim
0.6$ mag. Most of that excess is due to metal-poor main sequence
stars. Second, the models do not reproduce metal-sensitive optical
absorption features that arise mainly from red giant stars. Third, the
strength of a Ca II index sensitive to hot stars does not jibe with
the predicted number of A-type horizontal branch stars.  The number of
metal poor stars in galaxies is at least a factor of two less than
predicted by the Simple model, exactly similar to the ``G Dwarf
problem'' in the solar cylinder. Observations at larger radii in local
group galaxies indicate that the paucity of metal poor stars applies
globally, rather than only in the nuclei.

Because of the dominance of metal rich stars, primordial galaxies will
have a plentiful dust supply early in their star formation history,
and thus will probably have weak Ly$\alpha$ emission, as is apparently
observed.  We confirm that early-type galaxies cannot have been formed
exclusively from mergers of small all-stellar subsystems, a result
already established by dynamical simulations. The constraint of peaked
abundance distributions
will limit future chemical evolution models. It
will also make age estimates for the stellar populations in early type
galaxies and bulges more secure.

\end{abstract}
  

\section{Introduction} 

When forming a stellar system out of gas, supernovae explosions,
planetary nebulae, and quieter mass-loss enrich the remaining gas so
that succeeding generations of stars are more metal-rich than their ancestors. 
Uncomplicated analytic models are helpful for
exploring the basics of the chemical evolution problem, and the most uncomplicated of
them all is termed the ``Simple'' model, with capitalization intact
(\cite{s63}; \cite{ss72}; \cite{pp75}; capitalization as
preferred by Pagel, e.g. \cite{p89}). 
The Simple model
assumes (1) that the galaxy is represented by one zone, initially full
of metal-free gas, eventually full of stars, (2) that it is a closed
box with no inflow or outflow, (3) that enrichment occurs immediately
upon forming new stars (the Instantaneous Recycling Approximation), 
and (4) that the stars produce a constant yield 
($y=M_{\rm heavy}/M_{\rm H+He}$) of
heavy elements returned to the interstellar medium. 
The metallicity of
currently forming stars is set solely by the gas to total mass
fraction $\mu = M_{\rm gas}/M_{\rm stars + gas}$ by $Z_{\rm gas} = M_{\rm heavy}/M_{\rm
H+He} = 1 - \mu^y$ (\cite{ta71}; equation 16.15 in \cite{c95}). 
In a
completely stellar system, $\mu$ goes to zero, and the distribution of
stellar metal abundances is set by the yield $y$
\footnote{Although this formula gives $Z=1$ at $\mu=0,$ in practice this does 
not introduce a problem since the 
number of extreme-abundance stars involved is small.}.

Data from the solar vicinity has been compared to analytic models, and
the Simple model in particular (cf. \cite{r91} and references
therein). 
Even after correcting the purely local data to include stars
at high $Z$ in the solar cylinder, the Galactic data still displays a
paucity of metal-poor stars compared to the Simple model (\cite{r91},
\cite{c95}). 
This is referred to as the ``G Dwarf problem'' after the
unevolved tracer stars used for the abundance determinations. 
The G Dwarf problem can be alleviated in a variety of ways,
including gas infall, gas outflow, prompt initial enrichment,
consideration of a spatially inhomogeneous metal abundance pattern
(\cite{mhcm93}), or of a variable stellar yield. 
In addition, many
modelers now consider many-zone models, many track individual
elements (as opposed to one overall heavy-element enhancement), and
many drop the Instantaneous Recycling Approximation (e.g. the models
compared by \cite{t96}). 

In addition to the classical G dwarf problem, which is mostly
considered to apply to stars in the [Fe/H] $=-1.0$ to $-2.5$ range,
there is ``the flagrant scandal that we so rarely mention in public:
where is the First Generation?'' (\cite{k77}). Nearly primordial stars
with very small metal content ([Fe/H] $<-3.5$) are difficult to find
(\cite{bps92}), and a good question is whether they are present in the
numbers predicted by the Simple model.

Until recently, the solar vicinity was the only region in which
chemical evolution models could be applied.  The metal-poor dwarf
spheroidals have small spreads in metallicity based on the width of
the red giant branch (RGB) much like Galactic globulars, the vast
majority of which are tightly constrained to a single metallicity
throughout (\cite{st92}). \cite{r88} observed 88 K giants in the
Galactic bulge and derived a distribution of metal abundances that he
compared with the Simple model. In 1988 the distribution appeared to
match the Simple model fairly well, but subsequent recalibration of
the metallicity scale (\cite{mr94}) narrows the distribution somewhat
by pushing the very metal rich stars back toward solar
metallicity. Adjustments both for differences of RGB lifetime with
[Fe/H] and for the fact that metal-rich K giants are underrepresented
because they become M giants further narrow the abundance
distribution.  Recently, other galaxies have begun to be studied. A
relatively deep HST color-magnitude diagram of individual stars at
$1.8 \: R_e$ in local group elliptical M32 (\cite{getal96}) has been
decomposed into a metallicity distribution that is significantly more
peaked than the solar neighborhood, along with a field in the outer
disk of M31 which shows a distribution almost as narrow. \cite{rmn96}
have analysed HST images in the halo of M31, finding an unexpectedly
metal-rich environment poor in metal-deficient stars compared to our
own halo.

What about other places in the universe?  For distant galaxies and
dense regions like the nuclei of M31 and M32, integrated starlight
gives the only clues to the chemical composition currently available.
Here, we compare the integrated light of E/S0 galaxies and spiral
galaxy bulges, thought to contain old ($t > 2$ Gyr) passively evolving
stellar populations, with models, both empirical and theoretical.  We
show evidence for metallicity distributions in such systems that are
more strongly peaked than the Simple model allows in \S 2, and discuss
these findings in \S 3.

\section{Mid-UV Spectrum and Optical Absorption Features}

\begin{figure}	
\plotone{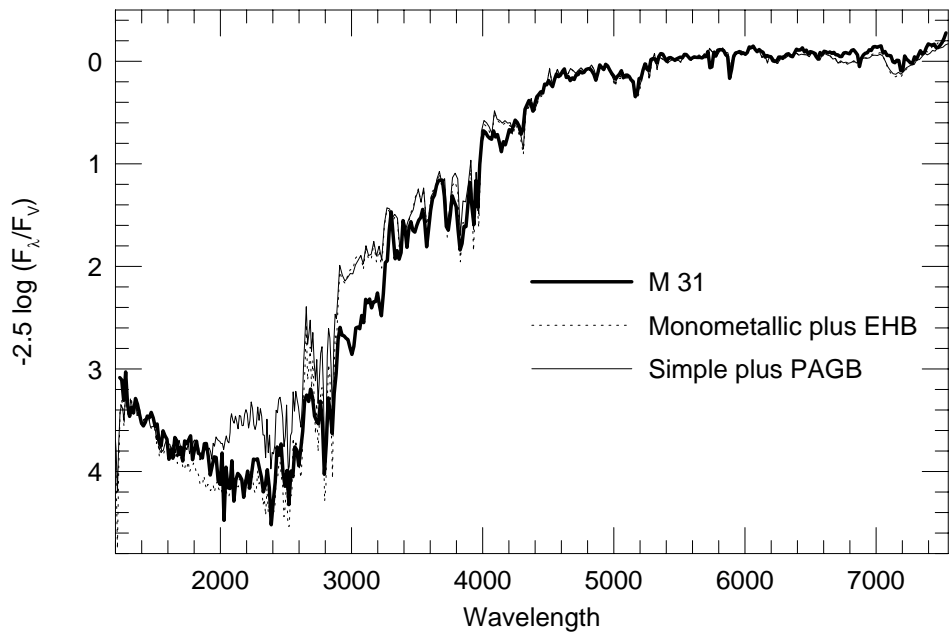}
\caption{\label{fit} A spectrum of M31 (D. Calzetti 1995, private
communication), normalized to the visual, is plotted as a bold line
with two theoretical models. The dotted line represents a model of a
single metallicity ([Fe/H] $=+0.05$), and age (12 Gyr). This model
incorporates an additional contribution of UV light from an EHB
(extreme horizontal branch) as modeled by \protect\cite{dor93} and a
PAGB (post-asymptotic giant branch) from the $0.565
\protect{M_{\sun}}$ sequence of \protect\cite{sc83}.  The thin solid
line represents the combination of models of different metallicities
in the proportions dictated by the Simple model, at age 17 Gyr and
stellar yield $Y={\rm log}(y/Z_\odot) =+0.3$.  The model metallicity
distribution was cut off above +0.5.}
\end{figure}

\placefigure{fit}

\cite{bcf94} and \cite{c96} have made population models of different
metallicities and noted that the addition of substantial ($\approx$
10\% by mass) portions of metal poor stars to an otherwise metal rich
population causes a discrepancy between theory and observation in the
ultraviolet spectrum. The problem is illustrated in Figure 1 for the
nucleus of M31 using two different models, one with a single
metallicity and one with a Simple model spread of metallicities. The
Simple model spectrum shows a mismatch around 2100\AA.  There are two
leading potential contributors to the flux in this region
(\cite{bbbfl}; \cite{dor95}; \cite{do96a}): (a) the hotter horizontal
branch (HB) stars that are often seen in Galactic globular clusters;
and (b) the flux from the metal-poor main-sequence (MS) turnoffs,
which affects the spectral energy distribution (SED) at progressively
shorter wavelengths at lower metallicity.  Metal-poor Galactic
globular clusters often contain HB stars hotter than $\sim 8000$ K in
large numbers, and these emit copiously in the mid-UV. They would thus
most likely be present if the composition distribution matched the
Simple model prediction. However, the process by which the HB stars'
location on the HR diagram is determined (mass loss on the red-giant
branch) is not at all understood, and thus conclusions arising from
the assumed behavior of the HB morphology with metallicity or age are
inherently unreliable.  However, by subtracting the maximum expected
contribution of hot HB star light from the models, we confirm that it
is the {\em MS turnoff light} that is mostly responsible for the
mismatch of the SED.

More quantitatively note that, in a 10-Gyr population with [Fe/H]
$\sim -1.5$ and an entirely blue HB, the turnoff light contributes
about 50\% of the flux at 2500~\AA\ (\cite{dor95}; \cite{do96b}).  At
this metallicity the main sequence contribution to the mid-UV flux is
about 2 mag brighter than that of solar metallicity.  An upper bound
the effect of the blue HB can be derived from the case of a
bimodal-metallicity population with metal-poor contribution similar to
that of the Simple model.  Using these rough estimates, and assuming
the metal-poor component only produces blue HB stars, a population
with 10\% of the stars with [Fe/H] $\le -1$ will have a blue HB
contribution $ < 0.3$ mag, whereas the discrepancy is $\sim 0.6$
mag. Estimates using the actual Simple model abundance distribution
for a solar yield and the Worthey models confirm this: only about
\onethird\ of the 2500~\AA\ flux is from blue HB stars, the rest from
the MS turnoff.

Figure~\ref{fit} shows the UV/Optical spectrum of M31 (D. Calzetti
1995, private communication), normalized to the $V$ band. The UV
spectrum (\wless{3300}) is from IUE (\cite{bbbfl}), while the optical
spectrum was taken with the IRS instrument at the Kitt Peak 0.9m
(\cite{mck95}).  The two models plotted are as follows: The dotted
line represents a model of a single metallicity ([Fe/H] $=+0.05$), and
age (12 Gyr), while the solid line is a Simple model, at age 17 Gyr
and stellar yield $Y={\rm log}(y/Z_\odot)=+0.3$.  The model
metallicity distribution was cut off above +0.5. Note that this model
is at the red extreme of what the Worthey (1994) models are capable of
producing.  The difficulty of finding a model red enough is another
(albeit weaker) argument that the Simple model distribution is not
appropriate for real galaxies. This model has additional contribution
from a very hot PAGB component fitted to the 1500\AA\ continuum: we
chose to add this type of hot component to supply the observed far-UV
flux as it has the smallest possible contribution around 2500~\AA.
Both theoretical spectra match the observation throughout the optical,
with stronger deviations toward the IR and UV. Some of this is
undoubtedly due to inaccurate line opacity in spectral ranges not yet
adequately charted.  Both the single metallicity model and the Simple
model fail in the spectral region {\wrange{3000}{3400},} where the
models appear too bright by $\sim 0.4$ mag.  The reason for this
discrepancy is unclear, but it also occurs in other galaxy spectra
(see \cite{do96a}).  If it is astrophysical, rather than due to
miscalibration of the IUE LWR spectra, then it may contain important
information about either stellar populations or model stellar
atmospheres.  The other notable feature is in the range
{\wrange{2000}{2600}:} the single-metallicity model is a fairly good
match, but the multiple-metallicity model predicts $\sim 0.6$ mag too
much flux. This factor of 1.7 is a {\em minimum} given the
conservative assumptions of the spectral fit.  Other reasonable models
will have worse mismatches.

\begin{figure}

\plotone{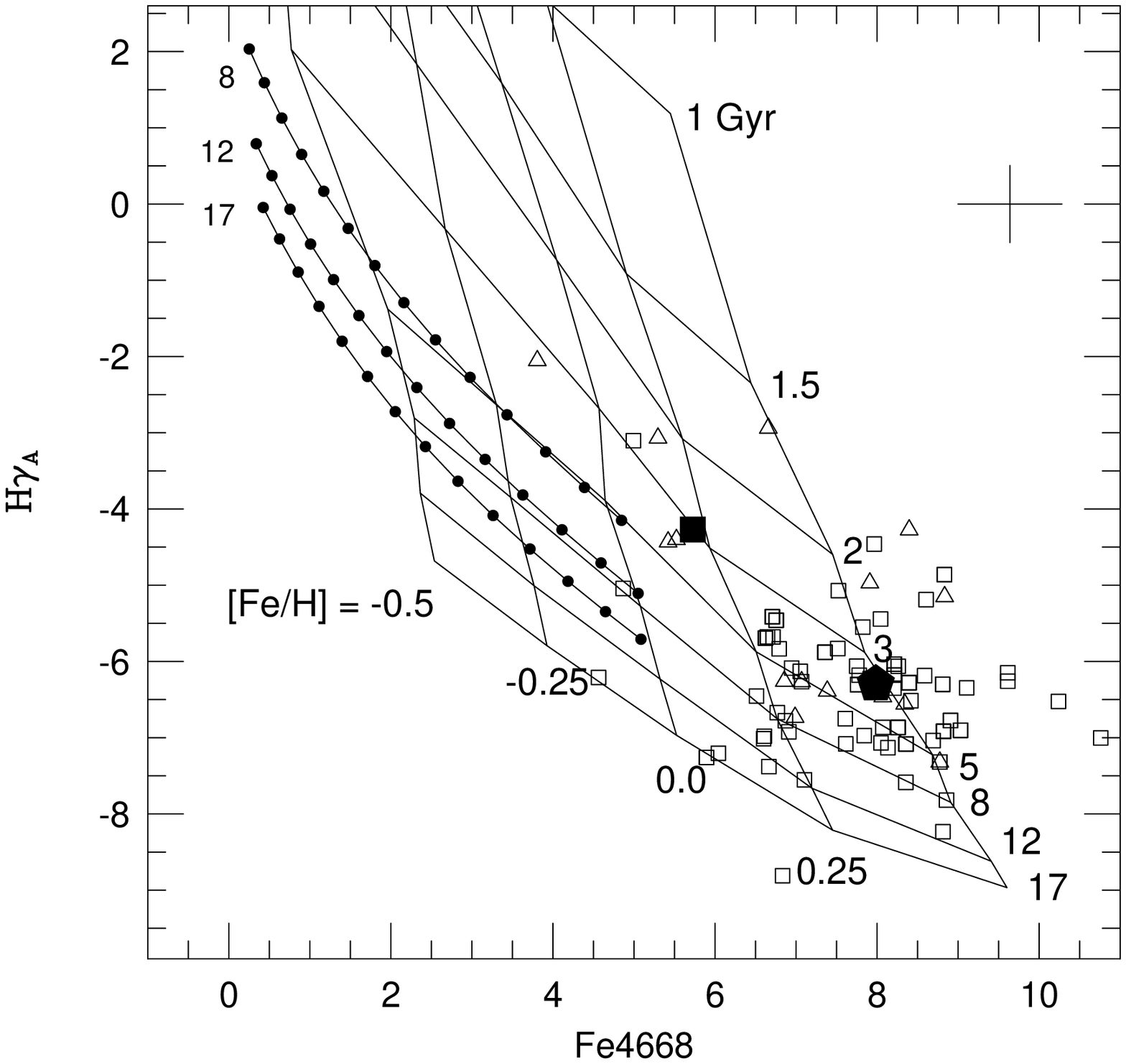}
\caption{\label{fig2} Galaxies and models are shown in an age-diagnostic
H$\gamma_A$ versus Fe4668 diagram
(see \protect\cite{wfgb94} and \protect\cite{jw95}). Nuclear galaxy data
are from \protect\cite{tetal96} and from yet unpublished
Balmer index data. Galaxies are a mixture of S0 (triangles) and
elliptical (squares) galaxies of different sizes and environments.
M31 appears as a solid pentagon and M32 as a solid square.
\protect\cite{w94} single-metallicity models are shown as a grid.
Fe4668 is primarily sensitive to metallicity, and values of
[Fe/H] are marked near the bottom of the grid. Ages in Gyr are marked
along the right side. Multiple-metallicity Simple models of ages 8,
12, and 17 Gyr are
illustrated as the 3 curves shifted to weak Fe4668 strength. 
The weakest models have yield $Y=-1$, and dots
appear along the curve at 0.1-dex intervals, ending with yield
$Y=+0.3$. }
\end{figure}

\placefigure{fig2}

It was difficult to find a Simple model red enough to match M31's
spectrum in Figure~\ref{fit}. Attempting to match a metallic
absorption feature that is particularly metallicity sensitive makes it
impossible to fit such a model, no matter how old or metal-rich.
Figure~\ref{fig2} shows the relatively age-sensitive H$\gamma_A$ index
against abundance-sensitive Fe4668 (\cite{wfgb94}; \cite{jw95}) for early-type
galaxies, with single-metallicity models and multi-metallicity models
that follow the Simple model abundance distribution for different
yields.
The Simple models lie at weak Fe4668, with yields running from $Y=-1$
at the weak end to $Y=+0.3$ with dots at 0.1 dex intervals.
The galaxy line strengths are seen to be well in excess of those
predicted by the Simple models, whereas the single-metallicity models
(solid grid lines) are able more easily to reproduce the galaxies'
line indices with abundance assumptions closer to expectation.
Shifting to a Simple model distribution adds a burden of
+0.3 dex of extra metal enhancement that the stars must have in order
to match the line strengths of single-metallicity models.
To match the galaxies clustered
at [Fe/H] $\approx +0.5$, the yield would have to be $+0.8$
($Z=0.13$), and we doubt that this is astrophysical.

The difficulty applies equally to E and S0 galaxies that span
the range in size and cluster/field environment, and to M31 as well.
The result is independent of which index or index combination one
chooses, as long as the more metal-sensitive features are used.
The result is also independent of which
sets of isochrones we use. The amount of metal-poor stars needs to be
reduced by at least a factor of two from the Simple model amount in
order for model metallic absorption features to match real galaxies,
but a factor of two reduction still requires uncomfortably large
metallicities (+0.5 to +0.7) for the bulk of the populations. A factor
of 4 reduction in metal poor stars gives more reasonable
metallicities, and is more consistent with the (\cite{getal96})
abundance distribution for M32.

\begin{figure}

\plotone{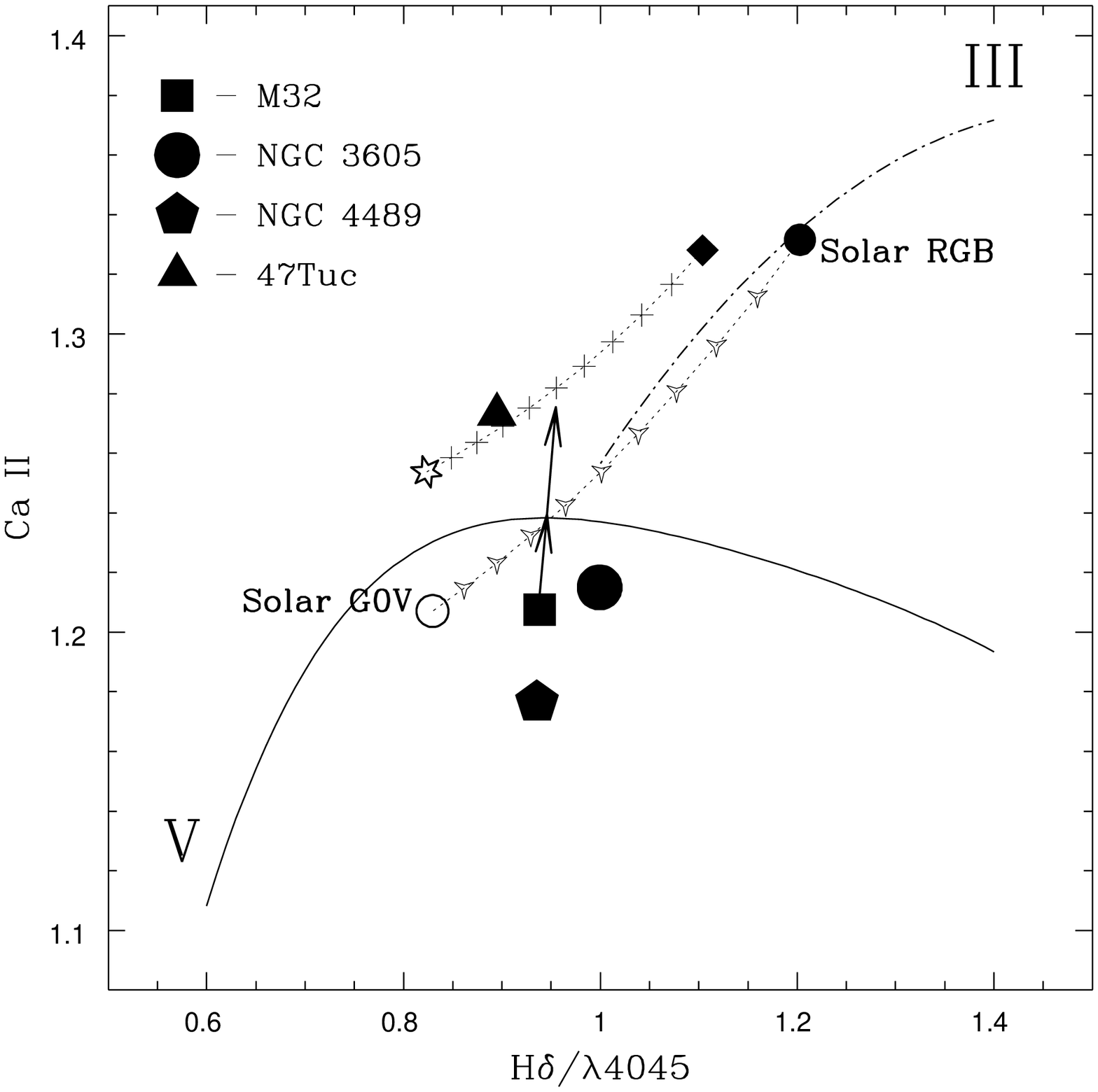}
\caption{\label{lewis} High resolution spectral indices for compact
ellipticals M32, NGC~4489, and NGC~3605 and Galactic globular
47~Tucanae.  CaII measures the ratio of Ca H + H$\epsilon$ and Ca K
line bottoms and is sensitive to the presence of stars hotter than F2.
H$\delta$/$\lambda$4045 is a spectral type indicator
(\protect\cite{ros94} and references therein). Observational index
errors are smaller than their symbols on the page.
Mean stellar sequences of dwarfs (luminosity class {\protect\sc v})
and giants (luminosity class {\sc iii}) of different temperatures and
metallicities are shown by the solid and dashed-dotted curve.  A mean
solar metallicity G0 dwarf (turnoff star) is labeled ``solar G0 {\sc
v},'' and a $\rm [Fe/H] = -0.5$ dwarf composite lies near 47 Tuc.  The
$\rm [Fe/H]=0$ (filled diamond) and --0.5 (filled circle) mean giants
are also plotted.  Dotted lines with cross ($\rm [Fe/H] = -0.5$) and
tricorn ($\rm [Fe/H] = 0$) symbols placed at each 10\% increment trace
different combinations of dwarf $+$ giant light around 4000 \AA.  The
arrows attached to M32 indicate the effect of subtracting from the
galaxy light contributions to the indices equivalent to 2\% and 3\%
blue horizontal branch (BHB) populations.  This is to be contrasted
with the 6\% blue HB populations predicted from the Simple model (see
text). }
\end{figure}

\placefigure{lewis}

Finally, Figure~\ref{lewis} shows three compact elliptical galaxies
(velocity dispersion less than 100 km s$^{-1}$)
from \cite{j96} in a diagram diagnostic of A stars (\cite{ros94}). There
are no synthesis models in this diagram, but plausible ``galaxies''
are constructed by combining observed giant and dwarf stellar spectra
at the appropriate metallicity. Hot stars lie at very low
values of CaII, and the arrows attached to M32 show how the index
shifts if small amounts of blue horizontal branch (BHB) light is subtracted. 
The vectors for the other galaxies are similar to those illustrated 
for M32.
A Simple model with yield chosen to match the
colors of these galaxies ($Y=0$) predicts 11\% {\em by mass} of the
population is in stars of [Fe/H] = --1 or less; populations that
should display BHBs if they are old.  The implied
fraction of light around 4000\AA\ (via Worthey models) that should come
from blue HB stars is about 6\%.  The observed metal-poor A star light
is 2\% for M32 and NGC~3605 and 3\% for NGC~4489,
about \onethird\ to \onehalf\ that predicted by the Simple model.
In a nutshell, the observed value of the Ca {\sc ii} index in the 
galaxies is too large for them 
to contain as many blue HB stars as the Simple model implies.

\section{Discussion}

The previous section presented arguments for a stronger-peaked
distribution for S0 and Elliptical galaxies of all sizes and
environments and the central regions of spiral M31. Studies of
individual stars in local group galaxies M31, M32, and the Milky Way
can extend this result to large radii.  Considering first the Galaxy,
we are mainly interested in assessing if the whole Galaxy, integrated
over every star, still fails to match the simple model. The reason for
this peculiar viewpoint is that standard dynamical collapse models
that include chemical evolution (e.g. \cite{l74}) predict that gas
enriches as it collapses, so that metal poor stars are stored
primarily at large radii, and the nuclei have high abundances with few
metal poor stars present. Abundance gradients, more metal poor
outwards, are a general feature of observed disks and spheroids. Is
there enough metal-poor material in the outskirts of galaxies to
approach the Simple model distribution if every star is included? For
our Galaxy, apparently not.  Looking at stars of low abundance, [Fe/H]
$<-1.5$, the solar neighborhood has virtually zero (\cite{r91};
\cite{p89}), and none have been found in the Bulge to date, either
(Rich 1988). The halo at the solar radius has a plentiful supply
(\cite{rn91}) but the metal-poor halo weighs only about $1\times 10^9
M_\odot$ (e.g. \cite{f96}) compared to about $60\times 10^9 M_\odot$
for the total mass of stars in the Galaxy, and therefore, via the Ryan
\& Norris (1991) halo star abundance distribution, about 1\% of the
total mass in the Galaxy is composed of stars of [Fe/H] $<-1.5$. This
is compared to about 3\% expected from the Simple model of solar
yield.

Interestingly, we can perform a similar back-of-the-envelope calculation
for stars between [Fe/H] $=-3.5$ and $-4.5$ by normalizing the numbers
of such very metal poor stars from Beers et al. (1992) to the Ryan \&
Norris (1991) distribution between [Fe/H] $=-2.5$ and $-3.5$. We
arrive at the conclusion that the Galaxy ought to have about $25\times
10^6 M_\odot$ of very metal poor stars, as compared to about
$20\times 10^6 M_\odot$ for the Simple model. This good agreement with
a global solar-yield Simple model means that the
number of extreme stars is
in excess of the number predicted by the modified Simple
model used in Ryan \& Norris. We would suggest that further work on
the abundance distribution of extremely metal poor stars might be
extremely interesting.

The situation in M31 shows signs of being even more extreme than our
own Galaxy in terms of lack of metal poor stars. An HST
color-magnitude diagram in the outskirts of the disk of M31
(\cite{getal96}) shows an abundance distribution extremely similar to
that of the solar neighborhood. The halo of M31 appears to have a
preponderance of metal rich stars, some apparently as metal-rich as
the sun, and few metal poor stars (Rich et al. 1996).  In M32 there is
as yet no information at very large radius, but at the the 65\%
light-enclosed radius ($\approx 1.8 R_e$), the abundance distribution
is significantly narrower than that of the solar neighborhood
(\cite{getal96}).

We therefore conclude that,
from small galaxies barely large enough to attain
solar metallicity to gigantic elliptical
galaxies, and from disks to spheroids regardless of field/cluster
environment, and from the nuclei alone to consideration of every star
in the galaxies there is no evidence for an
abundance distribution as broad as the Simple model predicts. 
Despite the small number of observations we cite
for spiral disks and bulges (two:
ours and M31's), it is quite likely that the G dwarf
problem is universal; that {\em nowhere in the universe does chemical
enrichment produce as many metal-poor stars as predicted by the Simple
model if the enrichment process proceeds to near-solar composition.}

This sweeping statement should provide a basic constraint for galaxy
evolution. Whatever processes are responsible for galaxy formation, by
the time $\sim$2\%--5\% of the final mass was assembled into stars the
95\%--98\% reservoir of gas was already quite metal-rich.  As galaxies
assembled, the periods in which metal-poor stars provided the bulk of
the stellar luminosity must have been very brief.

The universality of the G-dwarf problem
may be important in predicting what primeval
galaxies look like. Galaxies observed at high redshift will tend to be
the largest ones, already well into the process of forming
stars. Since the present-day abundance distribution is narrow, we
would expect high metallicities early in the history of the galaxies,
even if only a few percent of the eventual number of stars has
formed. High metallicities would mean a ready supply of dust, which is
now fingered as the likely cause of quenched Ly$\alpha$ in nearby HII
galaxies (\cite{p94}). High redshift ($z>3$) galaxies appear to be
very similar to local ones (\cite{sgrda96}) in terms of nearly
zero Ly$\alpha$ emission (this makes searching for primordial galaxies
by looking for Ly$\alpha$ emission seem unproductive.) A dust-free,
metal-poor phase of galaxy formation as discussed in Pritchet (1994)
would be considerably shorter than 30\% of the star-formation
timescale. Perhaps 5\% would be a better first guess, as that is the
upper limit on mass locked into metal-poor stars in present day galaxies.

About early-type galaxies we learn that
they cannot have formed primarily from
the merging of purely stellar satellite galaxies like dwarf
spheroidals. If this were the case, they would have many more metal
poor stars than they do. That pure-stellar mergers are unworkable
is known from dynamical simulations (e.g. \cite{v83}, \cite{h93}), which
show that the merging of purely stellar galaxies results in 
a large, flat core that is dissimilar from
real galaxies.

Beyond this, the narrower-than-Simple result will loosely constrain
numerical models of elliptical formation that include chemical
evolution (e.g. \cite{ay87}, Bressan et al. 1994, \cite{g96},  \cite{mt87}) by
eliminating those modeled abundance distributions that approach the
breadth of the Simple model. Plenty of room for variation will still
exist for these models.

The biggest benefit from knowing the basics of the metallicity
distribution of stars in early-type galaxies may be that it helps to
estimate ages for the component stellar populations. Figure~\ref{fig2}
shows an age diagnostic diagram in which the metallicity shifts are dramatic
depending on the breadth of the metallicity distribution, but the age
shifts are fairly mild. Knowing the approximate sharpness of the
abundance spread will make age estimates more secure in the future.

Why does the Simple model
fail universally? The simplest explanation would be that
chemical enrichment is predestined to happen extremely rapidly.  In
the context of the Simple model, the assumption to drop is the one of
constant stellar yield because on large spatial scales, considering every
member star, galaxies are closed boxes and major net infall or
outflow cannot be invoked.  The first idea for rapid enrichment that
comes to mind is that the IMF of metal-poor stars was
heavily skewed toward stars more massive than 1 $M_{\sun}$ so that
little mass was locked into unevolving stars and most of the enriched
gas remained available for future star formation. There is a
theoretical argument based on expected gas temperature suggesting that
this could be the case (e.g. \cite{af96}).  Since few low-mass stars
were formed, unevolved tracers would be rare.
The alternative explanation, that the
metal-poor stars are present but are stored at large radii, seems
ruled out in our Galaxy and M31, but more work is needed to be
certain.

\acknowledgements
Thanks are due to Daniela Calzetti, who provided the homogenized
spectrum of M31 in advance of publication.
This work was partially supported by NASA through grant
HF-1066.01-94A awarded by the Space Telescope Science Institute which
is operated by the Association of Universities for Research in
Astronomy, Inc., for NASA under contract NAS5-26555, and by 
NASA RTOP 188-41-51-03.

\end{document}